\documentclass[onecolumn, 11pt]{revtex4}
\usepackage{graphicx}
\usepackage{amsmath, amsthm, amssymb}
\newtheorem{theorem}{Theorem}

\newtheorem{lemma}{Lemma}

\newcommand{\wwto}{\stackrel{w}{\to}}

\newcommand{\supp}{\text{supp}  }
\providecommand{\norm}[1]{\lVert#1\rVert}
\providecommand{\abs}[1]{\lvert#1\rvert}
\newcounter{foo}

\begin{document}

\title{Bound States at Threshold. Many-particle case.}

\author{Dmitry K. Gridnev}
\email[Electronic address:] {gridnev|at|fias.uni-frankfurt.de}
\affiliation{FIAS, Ruth-Moufang Strasse 1, D--60438 Frankfurt am Main, Germany} 
\altaffiliation[On leave from:  ]{ Institute of Physics, St. Petersburg State University, Ulyanovskaya 1, 198504 Russia}
\begin{abstract}
We consider the phenomenon of eigenvalue absorption for a many body Hamiltonian, which depends on a parameter.
The conditions on pair potentials, which guarantee that the eigenvalues
approaching the bottom of the continuous spectrum become absorbed as the parameter approaches a critical value, are derived. We also discuss the behavior of bound states' wave functions when the corresponding levels approach the bottom of the continuous spectrum. The results have applications in atomic and molecular physics. An application to the stability problem of three Coulomb charges is presented.
\end{abstract}

\maketitle


\section{Introduction}\label{sec:sec1}

In this paper we consider a Hamiltonian $H(Z)$ depending on a parameter $Z$.
For $Z$ in the neighborhood of some critical value $Z_{cr}$ the system is supposed to have a bound state
$\psi (Z) \in L^2 $ with the energy $E(Z)$ and when $Z \to Z_{cr}$
the energy approaches the bottom of the continuous spectrum $E(Z) \to
E_{thr}$. The question is then whether $H(Z_{cr})$ has a bound state exactly at the bottom of the continuous spectrum. If the answer is yes then one speaks of eigenvalue absorption.
The two-body case with $Z$ being the coupling constant of the interaction is well--studied \cite{we,gest,newton}. In particular, it is known \cite{we,gest} that the eigenvalue absorption takes place for the potentials with a positive part falling off slower than $r^{-2}$. To our knowledge, there is only one result of this sort in the many body case, namely, the brilliant proof by Thomas and Maria Hoffmann-Ostenhof together with Barry Simon \cite{ostenhof} that a two-electron atom with an infinitely heavy nucleus has a bound state at threshold, when the nuclear charge becomes critical. The role of the parameter $Z$ in their proof is played by the nuclear charge. Our aim here is to investigate the general many body case. The theorems presented in the paper generalize, in particular, the result in \cite{ostenhof} to the case of finite non--equal masses.

Let us say a few words on the physics of eigenvalue absorption. Our view is that the main physical effect of this phenomenon lies not in a  mere presence of a bound state exactly at threshold but rather in the behavior of wave functions of those energy levels, which approach the bottom of continuum. For example, from absence of a bound state at threshold it follows that $\psi(Z)$ fully spreads for $Z \to Z_{cr}$ (see Sec.~\ref{sec:sec2}), {\it i.e.} the probability to find all particles in any fixed bounded region of space goes to zero. This, in turn, means that the size of the system goes to infinity. Such physical effect when a size of a bound system drastically increases near threshold can be found in neutron halos, helium dimer, Efimov states, Rydberg states etc., for
discussion see Refs.~ \cite{we,fedorov,zhukov,efimov,hansen0}.
In this paper the connection of eigenvalue absorption to spreading or non--spreading of bound states is used as a key method in the proof. In other words, we prove the fact of eigenvalue absorption by demonstrating that the corresponding bound state wave functions do not spread. In the following we demonstrate that the many body case is similar to the two body problem \cite{we} in the sense that a long-tailed repulsive interaction between possible decay products prevents spreading and forces an $L^2$ bound state at threshold. In nuclear physics, the non--spreading of bound states in the presence of a positive Coulomb tail in the interaction between possible decay products explains why no proton halos are found \cite{hansen}.

The article is organized as follows. In Sec.~\ref{sec:sec2} we discuss
the connection between spreading and eigenvalue absorption. The basic notations are also introduced here. In Sec.~\ref{sec:sec3} we discuss the bounds on two--particle Green's functions from \cite{we}. In Sec.~\ref{sec:sec3} we consider a system with no bound subsystems and prove that if pair interactions are repulsive and fall off not faster than $r^{-2}$ then at the critical point there is a zero energy bound state. In Sec.~\ref{sec:sec4}
we provide similar results for the case where the dissociation threshold
is formed by two clusters. In the last two sections we discuss an application to the problem of three Coulomb charges, which extends the result in \cite{ostenhof}, and draw conclusions.

\section{Spreading and Bound States at the Threshold}
\label{sec:sec2}
To make the bound states approach the threshold we make the interactions in the Hamiltonian depend on a
parameter $Z \in \mathbb{R}^p$ (we have chosen $\mathbb{R}^p$ for the sake of clarity, but in fact, the nature of the parameter space does not play a role). Throughout this paper under the \emph{Hamiltonian with a parameter sequence} we shall mean the pair $(H(Z), \mathcal{Z})$, where $Z \in \mathcal{Z} \subset \mathbb{R}^p$ and the set $\mathcal{Z}$ consists of a given sequence of parameter values $\{Z_k\}_{k=1}^\infty$ converging to some critical
value $Z_k \to Z_{cr}$ and the limit point itself, that is $\mathcal{Z} := \{Z_k\} \cup Z_{cr}$. The Hamiltonian $H(Z)$ describes the system of $N$ particles
\begin{gather}
    H(Z) = H_0 + V(Z, x) \label{xc31} \\
V(Z, x) =  \sum_{1 \leq i<j \leq N} V_{ij} (Z; x_i - x_j ) \label{:xc31},
\end{gather}
where $H_0$ is the kinetic energy operator with the center of mass
removed, $x_i \in \mathbb{R}^3$ denote particles' position
vectors and $x \in \mathbb{R}^{3N-3}$ denotes the full set of relative coordinates. The pair potentials are subdued to the following
restrictions

\begin{list}{R\arabic{foo}}
{\usecounter{foo}
    \setlength{\rightmargin}{\leftmargin}}
\item $|V_{ij} (Z; y )| \leq F (y)$ for all $Z \in \mathcal{Z}$, where $F(y)$ is such that
$\chi_{\{ |y| \leq d_0 \}}F (y) \in L^2
(\mathbb{R}^3 )$ and  $\chi_{\{ |y| \geq d_0 \} }F (y) \in
L^{\infty}_{\infty} (\mathbb{R}^3 )$ and $d_0$ is some positive constant.

\item $\forall f(x) \in C^\infty_0 (\mathbb{R}^{3N-3} )\colon  \lim_{Z_k \to Z_{cr}} \bigl\| \bigl[ V(Z_k) - V(Z_{cr}) \bigr] f \bigr\| =
0$, where $\{Z_k\} = \mathcal{Z}/Z_{cr}$.
\end{list}
In R1 the symbol $L^{\infty}_{\infty} $ denotes bounded Borel
functions, which go to zero at infinity.  For instance, $F (y) $
could be continuous apart from some square--integrable
singularities and falling off at infinity. With these
restrictions on the potentials $H(Z)$ is self-adjoint on $D(H_0 ) \subset L^2 (\mathbb{R}^{3N-3})$
 \cite{reed,teschl}.

By $E_{thr} (Z)$ we shall denote the bottom of the continuous spectrum of $H(Z)$, that is
\begin{equation}\label{sigmaess}
  E_{thr} (Z) := \inf \sigma_{ess} (H(Z))
\end{equation}

The set of requirements on the system continues as follows
\begin{list}{R\arabic{foo}}
{\usecounter{foo}
    \setlength{\rightmargin}{\leftmargin}}
\setcounter{foo}{2}
\item
for all $ Z_k \in \mathcal{Z}/Z_{cr}$ there are $E(Z_k) \in \mathbb{R}, \psi(Z_k) \in D(H_0)$ such that $H(Z_k) \psi(Z_k) = E(Z_k) \psi(Z_k)$, where $\| \psi(Z_k) \| = 1$ and $E(Z_k) < E_{thr}(Z_k)$.

\item
$\lim_{Z_k \to Z_{cr}} E(Z_k) = \lim_{Z_k \to Z_{cr}} E_{thr}(Z_k) = E_{thr} (Z_{cr})$, where $\{Z_k\} = \mathcal{Z}/Z_{cr}$.
\end{list}

The requirements R3-4 say that for all members of the parameter sequence the system has a level below the continuum and for $Z \to Z_{cr}$ the energy of this level approaches the bottom of the continuous
spectrum.

In the proofs we shall extensively deal with the so-called spreading sequences. The term is borrowed from \cite{zhislin}, where Zhislin used the idea of spreading sequences in his proof of the ``atomic'' version of the HVZ theorem. We shall say that the sequence of functions $f_n (x) \in L^2
(\mathbb{R}^n)$ \emph{does not spread} if for any $\varepsilon >0$ there exist $R , N>0$ such that $\|  \chi_{\{x||x| > R\}} f_n \| <  \varepsilon$ for all $n > N$. (From now and on $\chi_{A}$ denotes the
characteristic function of the set $A$). Otherwise the sequence is
called \emph{spreading}. The definition merely says that for a
non-spreading sequence $f_n$ no parts of the support of $f_n$
escape to infinity. We shall say that the sequence $f_n$
\emph{fully spreads} if the whole support of $f_n$ escapes to
infinity, \emph{i.e.} $\forall R\colon \lim_{n \to \infty} \|
\chi_{\{x||x| \leq R\}} f_n \| = 0$. Physically speaking, if the
sequence of wave functions fully spreads it means that the
probability to find all particles in any fixed bounded region of space
goes to zero. In the rest of this section we shall prove a number of technical results, which are in the spirit of \cite{simon} and serve as machinery in dealing with spreading sequences of wave functions. The following lemma and theorem give us the conditions,
which guarantee the non-spreading of sequences.

\begin{lemma}\label{rsd1}
Suppose that the sequence of functions $f_n \in L^2 (\mathbb{R}^n)$ is uniformly norm-bounded and $|f_n|$ is non-decreasing $|f_n | \leq |f_{n+1}|$. Then $f_n$ does not spread.
\end{lemma}
\begin{proof}
Let us assume by contradiction that $f_n$ spreads. Then there must
exist a positive constant $a > 0$ such that
$\limsup_{n \to \infty}\|\chi_{\{x||x| \geq R\}} f_n \| >
a$ for all $R>0$. Let us fix $n$ and choose $R$ so that $\| \chi_{\{x||x| \geq
R\}} f_n \|^2 < a^2 /4$. Because the sequence $f_n$ spreads we can
find $n'
> n$ such that $\| \chi_{\{x||x| > R\}} f_{n'} \|^2 > a^2 /2$.
Using that $|f_n |$ is non--decreasing we obtain
\begin{gather}
    \| f_{n'} \|^2 = \| \chi_{\{x||x| \leq R\}} f_{n'} \|^2 +  \|  \chi_{\{x||x| > R\}} f_{n'} \|^2 \geq  \| \chi_{\{x||x| \leq R\}} f_{n} \|^2 +  \|  \chi_{\{x||x| > R\}} f_{n'} \|^2 = \label{0.02}\\
\| f_{n} \|^2 -  \| \chi_{\{x||x| > R\}} f_{n} \|^2 + \|
\chi_{\{x||x| > R\}} f_{n'} \|^2 \geq     \| f_{n} \|^2  +
\frac{a^2}4 \label{0.03}
\end{gather}
Eqs. (\ref{0.02})-(\ref{0.03}) tell us that for any $f_n$ there
exists such $f_{n'}$ with $n' >n$ that $\| f_{n'} \|^2 \geq \| f_n
\|^2  + a^2 /4$. But this contradicts $f_n$ being a
norm-bounded sequence, hence, $f_n$ does not spread.  \end{proof}

This result can be made stronger.
\begin{theorem}\label{rsd2}
Suppose that a sequence $f_n \in L^2 (\mathbb{R}^n)$ satisfies the following inequality
\begin{equation}\label{0.01}
    |f_n| \leq  g_n  + | h_n |
\end{equation}
where $g_n , h_n \in L^2 (\mathbb{R}^n)$, $\| h_n \| $ are uniformly bounded and $g_n$ converges in norm. Additionally, suppose that the sequence $|h_n |$ has the following property: from any subsequence $| h_{n_k}|$ one can choose a sub/subsequence $| h_{n_{k_s}}|$, which is non--decreasing $|h_{n_{k_s}} | \leq |h_{n_{k_{s+1}}} |$. Then the sequence $f_n$ does not spread.
\end{theorem}
\begin{proof} Evidently, if a non-negative sequence is dominated by a non-spreading
sequence then it does not spread. It is also obvious that the sum
of two non-spreading sequences does not spread as well. The sequence $g_n$
does not spread because it converges in norm and to prove the lemma
we need only to show that the sequence $|h_n|$ does not spread. Then by (\ref{0.01}) $|f_n|$ is dominated by a sum of two non-spreading sequences and hence does not spread as well.

Let us assume by contradiction that $|h_n|$ spreads, which means
that $\limsup_{n \to \infty} \|\chi_{\{x||x|
\geq R\}} h_n \| >  a$ for all $R$, where $a > 0$ is a constant.
This means that for $k = 1,2,\ldots$ we can extract a subsequence
$h_{n_k}$ that satisfies $\| \chi_{\{x| |x| \geq k \}} h_{n_k} \|
> a$.

On one hand, it is easy to see that every subsequence of $|h_{n_k}|$ spreads. On the other hand, by condition of the theorem $|h_{n_k}|$ contains a subsequence, which is non--decreasing and uniformly bounded and thus cannot spread by Lemma~\ref{rsd1}. This is the contradiction and, hence, $|h_n|$ does not spread.  \end{proof}

An important consequence of restriction R1 on pair potentials is given by the following lemma.
We shall call the sequence $f_n \in D(H_0 )$ uniformly $H_0$-bounded if the
sequence $H_0 f_n$ is uniformly norm-bounded.

\begin{lemma}\label{forg}
Let $(H(Z), \mathcal{Z})$ be a Hamiltonian with a parameter sequence satisfying R1-4. Then the sequence $\psi(Z_k)$  defined in R3 is uniformly $H_0$-bounded.
\end{lemma}

\begin{proof}
The statement represents a
well-known fact, see f.e. \cite{zhislin} but for
completeness we give the proof right here.
By contradiction, let us assume that for $Z_k \to Z_{cr}$ we have $\| H_0 \psi (Z_k)\| \to \infty$.
By the Shr\"odinger equation $H_0 \psi (Z_k)= -V(Z_k)\psi (Z_k) + E(Z_k)
\psi(Z_k)$. Because $E(Z_k)$ are uniformly bounded and $\psi(Z_k)$ are normalized
we obtain the bound $\| H_0 \psi (Z_k) \| \leq \| V(Z_k
)\psi (Z_k) \| + O(1)$. From this inequality $\| H_0 \psi (Z_k)\| \to
\infty$ would mean $\norm{V(Z_k)\psi(Z_k)} \to \infty$ as well.

R1 tells us that the pair potentials $V_{ij}$ are bounded by $F_{ij}$, where for a shorter notation we denote $F_{ij} := F(x_i - x_j)$. Using
that as an operator $F_{ij}$ is $H_0$ bounded \cite{reed}
with a relative bound 0 we obtain the chain of inequalities
\begin{gather}
    \|V(Z_k) \psi (Z_k)\| = \bigl\| \sum_{i<j} V_{ij} (Z_k; x_i - x_j )  \psi (Z_k) \bigr\|
    \leq \frac{N(N-1)}2 \bigl\| F_{ij}  \psi (Z_k) \bigr\| \leq \label{kuka}\\
a \|H_0 \psi (Z_k)\| + b \leq
    a \| V(Z_k)\psi (Z_k)\| + O(1) \label{kuka:1}
\end{gather}
where $a, b >0$ are constants independent of $Z$ and $a$ can be
chosen as small as pleased. Taking, for example, $a = 1/2$ and
dividing both sides of the inequality (\ref{kuka})--(\ref{kuka:1}) by $\| V(Z_k)
\psi(Z_k) \|$ we find that the assumption $\| V(Z_k) \psi(Z_k) \| \to
\infty$, respectively $\| H_0 \psi (Z_k)\| \to \infty$ is false.
 \end{proof}

The following theorem illustrates the connection between non-spreading and bound states at the threshold.

\begin{theorem}\label{th:1}
Let $(H(Z), \mathcal{Z})$ be a Hamiltonian with a parameter sequence satisfying R1-4. If the sequence  $\psi (Z_k)$ defined in R3 does not fully spread then $H(Z_{cr})$ has a  bound state at the threshold
\begin{equation}\label{xc11}
H (Z_{cr}) \psi_0 = E_{thr}(Z_{cr}) \psi_0 ,
\end{equation}
where $\psi_0 \in D(H_0) \subset L^2 (\mathbb{R}^{3N-3})$.
\end{theorem}
Before we start with the proof we shall need a couple of technical Lemmas, which would be of use in the following sections as well.

\begin{lemma}\label{kuksi}
Let $f_n \in D(H_0)$ be a uniformly $H_0$-bounded sequence, which converges weakly $f_n \wwto f_0$. Then (a) $f_0 \in D(H_0)$; (b) if the operator $A$ is relatively $H_0$ compact then $\| A ( f_n - f_0 ) \| \to 0$.
\end{lemma}
\begin{proof}
First, let us prove that the sequence $H_0 f_n $ is weakly convergent. A proof by contradiction. By condition of the lemma $H_0 f_n $ is uniformly norm--bounded. Then if the sequence $H_0 f_n $ does not converge weakly there must exist at least two weak limit points, {\it i.e.} there exist two sequences $f'_k, f''_k$, which are subsequences of $f_n$ and for which $H_0 f'_k \wwto \phi_1$ and $H_0 f''_k \wwto \phi_2$, where $\phi_{1,2} \in L^2$ and $\phi_1 \neq \phi_2$. On one hand, because $\phi_1 \neq \phi_2$ and $D(H_0)$ is dense in $L^2$ there is $g \in D(H_0)$ such that $(\phi_1 - \phi_2 , g) \neq 0$. On the other hand, using that $f'_k \wwto f_0$ and  $f''_k \wwto f_0$ we get
\begin{equation}\label{:labi:}
    (\phi_1 - \phi_2 , g) = \lim_{k \to \infty} \left[ \bigl( H_0 (f'_k - f''_k), g\bigr)\right] =
    \lim_{k \to \infty} \left[ \bigl( (f'_k - f''_k), H_0 g\bigr)\right] = 0,
\end{equation}
a contradiction. Hence, $H_0 f_n \wwto G$, where $G \in L^2$. $\forall f \in D(H_0)$ by self-adjointness of $H_0$ we obtain $(H_0 f, f_0) = \lim_{n \to \infty} (H_0 f, f_n) = (f, G)$. Thus $f_0 \in D(H_0)$ and $G = H_0 f_0$, which proves (a).

To prove (b) note that $(H_0 + 1)(f_n - f_0 ) \wwto 0$. Using that compact operators acting on weakly convergent sequences make them converge in norm we get
\begin{equation}\label{fck}
A(f_n - f_0 )  = A(H_0 + 1)^{-1} (H_0 + 1) (f_n - f_0 )\to 0
\end{equation}
since $A(H_0 + 1)^{-1}$ is compact by condition of the lemma.  \end{proof}

In a different form the statement (a) of Lemma~\ref{kuksi} can be found in \cite{zhislin}. Let us remark that in this Lemma neither the nature of the Hilbert space nor the nature of the operator $H_0$ play a role, the only thing that matters is the self--adjointness of $H_0$.

\begin{lemma}\label{rsd3}
Let $f_n \in D(H_0)$ be an $H_0$-bounded sequence of functions, which converges weakly $f_n \wwto \phi_0 $. Then (a) if $f_n$ does not spread then $f_n \to \phi_0$ in norm; (b) if $f_n$ does not fully spread then $\phi_0 \neq 0$.
\end{lemma}
\begin{proof}
Let us start with (a).
Because $f_n$ does not spread it is enough to show that $\forall R$ $\| \chi_{\{x| |x| \leq R\}} (f_n - \phi_0) \| \to 0$ in norm. This follows if we apply Lemma~\ref{kuksi} and use that $\chi_{\{x| |x| \leq R\}}$ is relatively $H_0$ compact \cite{reed,teschl}.

Let us prove (b). Assume by contradiction that $f_n \wwto 0$. Using the arguments from (a) we get that $\forall R$ $\| \chi_{\{x| |x| \leq R\}} f_n \| \to 0$. But this would mean that $f_n$ fully spreads contrary to the condition of the Lemma.  \end{proof}

\begin{proof}[Proof of Theorem~\ref{th:1}]
Because $\psi(Z_k)$ does not fully spread there are $a, R > 0$ and a
subsequence $Z_n \in \mathcal{Z}/Z_{cr}$, $Z_n \to Z_{cr}$ such that $\| \chi_{\{x| |x| < R\}} \psi(Z_n) \| > a$ for all $n$. From this subsequence by the
Banach-Alaoglu theorem we choose a weakly convergent sub/subsequence
(for which for economy of notation we keep the notation $\psi(Z_n)$) $\psi(Z_n) \wwto
\psi_0$, where $\psi_0$ is the weak limit point and $\psi_0 \in D(H_0)$ by Lemma~\ref{kuksi}. Then the
sub/subsequence $\psi(Z_n)$ does not fully spread and is weakly
convergent, hence,  by Lemma~5 $\psi_0 \neq 0$. For any $f \in
C_{0}^\infty$ we have
\begin{gather}
    \Bigl([H(Z_{cr}) - E_{thr}(Z_{cr})]f , \psi_0 \Bigr) = \lim_{Z_n \to Z_{cr}} \Bigl( [H(Z_{cr}) -E_{thr}(Z_n)]f , \psi (Z_n) \Bigr) = \\
\lim_{Z_n \to Z_{cr}} \Bigl( \bigl[ H(Z_n) - (V(Z_n)-V(Z_{cr}) )  -E_{thr}(Z_n) \bigr] f ,\psi (Z_n) \Bigr)  = \\
\lim_{Z_n \to Z_{cr}} \Bigl\{ \bigl[E(Z_n) -E_{thr}(Z_n) \bigr] \Bigl( f ,\psi (Z_n) \Bigr)  -
\Bigl([V(Z_n)-V(Z_{cr}) ]f ,\psi (Z_n)\Bigr) \Bigr\} = 0,
\end{gather}
where in the last equation we have used R2,4.
Summarizing, for all $f \in C_{0}^\infty$ we have
\begin{equation}\label{xc12}
    \left(\bigl[H(Z_{cr}) - E_{thr}(Z_{cr})\bigr]f , \psi_0 \right) =  \left(f , \bigl[H(Z_{cr}) - E_{thr}(Z_{cr})\bigr]\psi_0 \right) = 0,
\end{equation}
meaning that Eq.~(\ref{xc11}) holds.  \end{proof}

Notice, that in the proof we had to consider a weakly converging subsequence of
bound states. The following Lemma is useful in dealing with general
sequences of bound states.
\begin{lemma}\label{rsd5}
Let $f_n \in L^2 (\mathbb{R}^n)$ be a normalized sequence of functions, with
the property that every weakly converging subsequence converges
also in norm. Then $f_n$ does not spread.
\end{lemma}
\begin{proof}
By contradiction, let us assume that $f_n$ spreads. In the proof
of Lemma~2 we have shown that for a spreading sequence $f_n$ it is
possible to extract a subsequence $g_k = f_{n_k}$ with the
property $\| \chi_{\{x| |x| \geq k \}} g_k \| > a$, where $a > 0$
is some constant. On one hand, it is easy to see that $g_k $ with
this property does not have any subsequences that converge in
norm. On the other hand, by the Banach-Alaoglu theorem $g_k$ has
at least one  weakly converging subsequence. This must also
converge in norm as a subsequence of $f_n$ and there is a
contradiction. Hence, $f_n$ does not spread. \end{proof}

Later we would show that for certain potentials the sequences of
bound states $\psi(Z_k)$ have exactly this
property:  all weakly converging subsequences converge in norm. By Lemma~\ref{rsd5} this would mean that
the whole sequence of bound states $\psi(Z_k)$ does not spread.

\section{Two--Particle Results Revisited}\label{sec:sec3}

In this section we return to the results of \cite{we} and rewrite the bounds on the Green's function of two particles in the context required for the present paper. Let us set $N=2$ and let $G_k (x,y)$ denote the kernel of the integral operator
\begin{equation}
G_k = \Bigl[ p^2 + \frac{3 + \delta}{4 |x|^2} \chi_{\{
x | \: |x| \geq n \} } + k^2 \Bigr]^{-1} ,
\end{equation}
where $x = x_2 - x_1$ is the relative coordinate, $p$ is the conjugate momentum with respect to $x$ and $k>0$.

In \cite{we} the proof of eigenvalue absorption in the two particle case is based on the following pointwise upper bound on the
Green's function $G_k (x,y)$, which after combining Eqs. (15),
(18) in \cite{we} has the form
\begin{equation}\label{remni}
    G_k (x,y) \chi_{\{ y| \: |y| \leq n\} } \leq \frac {\chi_{\{ y| \: |y| \leq n\} }}{4\pi |x-y|} \times  \left\{ \begin{array}{ll}
    1 - \tilde R^{-1}_0 \tilde a (\tilde a +1)^{-1} |x-y|  & \quad \textrm{if $|x-y | \leq \tilde R_0 $} \\
    \tilde R_{0}^{\tilde a} (1+\tilde a )^{-1} |x-y|^{-\tilde a} & \quad \textrm{if $ |x-y| \geq \tilde R_0$} \\
    \end{array}
    \right.
\end{equation}
where $\tilde a (y), \tilde R_0 (y)$ are real--valued
functions on $\mathbb{R}^3$ satisfying the inequalities (see Eqs.
(12)--(13) in \cite{we})
\begin{gather}
    \tilde R_0 \geq |y|+n \label{remni2a} \\
\tilde a(\tilde a  +1) \leq \frac{3+\delta }4 \frac{\tilde R^2_0
}{(\tilde R_0 + |y|)^2} \label{remni2b}
\end{gather}
It makes sense to simplify the inequality (\ref{remni}). Using that in
(\ref{remni}) only the values $|y| \leq n$ matter we can fix $\tilde a$ and
$\tilde R_0$ in the following way
\begin{gather}
\tilde a = \frac 12 + \frac{\min(1,\delta)}{20} \label{hungr1} \\
\tilde R_0 = \frac{20}{\min(1,\delta)}  n \label{hungr2}
\end{gather}
By the direct calculation one can check that the choice (\ref{hungr1})--(\ref{hungr2})
satisfies the inequalities (\ref{remni2a})--(\ref{remni2b}).
It is only important that $\tilde a
> 1/2$ and $\tilde a$ does not depend on $n$. No we can rewrite the inequality (\ref{remni}) as follows
\begin{equation}\label{remnix}
    G_k (x,y) \chi_{\{y| \:  |y| \leq n\} } \leq \frac {\chi_{\{y| \:  |y| \leq n\} }}{4\pi |x-y|} \times  \left\{ \begin{array}{ll}
    1   & \quad \textrm{if $|x-y | \leq \tilde R_0 $} \\
    C_\delta n^{\tilde a} |x-y|^{-\tilde a} & \quad \textrm{if $ |x-y| \geq \tilde R_0$} \\
    \end{array}
    \right.  ,
\end{equation}
where $\tilde a$ and $\tilde R_0$ are defined through (\ref{hungr1})--(\ref{hungr2}) and $C_\delta $ is a constant, which depends only on $\delta$. As one can see, the upper bound (\ref{remnix}) does not depend on $k$. Estimating through (\ref{remnix}) the Hilbert--Schmidt norm of the operator $G_k \chi_{\{ x| \: |x| \leq n\} } $  (which  is the product of the resolvent with the operator of multiplication by a characteristic function) tells us that $G_k \chi_{\{ x| \: |x| \leq n\} } $ is a Hilbert--Schmidt operator and its Hilbert--Schmidt norm is uniformly bounded for all $k> 0$. Generally, the operator $G_k F(x) \chi_{\{ x| \: |x| \leq n\} } $ is a Hilbert--Schmidt operator when $F \in L_{loc}^2 (\mathbb{R}^3)$. Of course, $G_k$ by itself is neither Hilbert--Schmidt nor it is uniformly bounded for $k\to 0$ but through the cut off by the characteristic function it gains both of these properties.

Below we demonstrate that to make $G_k$ uniformly bounded for $k \to 0$ it is sufficient to multiply it by a polynomially decaying function instead of a cut off through $\chi_{\{ x| \: |x| \leq n\} }$.

\begin{lemma} \label{expon}
Suppose $f(x) \in L^2_{loc} (\mathbb{R}^3)$ and there are $\varepsilon, m >0 $ such that $\sup_{|x| \geq m} \left( |x|^{3+\varepsilon} |f(x)|\right) < + \infty$. Then
\begin{equation}\label{k1}
\sup_{k} \| G_k f(x)\|_2  < + \infty,
\end{equation}
where $\| \cdot \|_2$ stands for the Hilbert--Schmidt norm of an operator.
\end{lemma}
\begin{proof}
For a shorter notation let us denote $\chi_n := \chi_{\{x|\: |x| \leq n\}}$
In view of (\ref{remnix}) and (\ref{hungr1})--(\ref{hungr2}) we have
\begin{gather}
    \left( \| G_k \chi_n \|_2 \right)^2 = \int_{|y| \leq n } dx dy |G_k
(x,y)|^2 \leq \int_{|y| \leq n } dy \int_{|x-y| \leq \tilde R_0 } dx
\frac 1{|x-y|^2} + \label{r1}\\
C^2_\delta n^{2\tilde a} \int_{|y| \leq n } dy \int_{|x-y| \geq  \tilde R_0 }
dx \frac 1{|x-y|^{2(\tilde a + 1)}} \label{r2}
\end{gather}
where we have used that $(4\pi)^{-2} < 1$. Note, that the second integral on the rhs converges because $\tilde a > 1/2$. Through the change of variables, namely
$z_1 = y$ and $z_2 = x-y$, the integrals on the rhs in (\ref{r2}) can be calculated explicitly. The direct calculation tells us that both integrals on the rhs of (\ref{r2}) are proportional to $n^4$. Hence, we can conclude that there exists a constant $b > 0$ depending only on $\delta$ such that
\begin{equation}\label{a1a}
\norm{  G_k \chi_n }_2 \leq b n^2 ,
\end{equation}
We can rewrite the Hilbert--Schmidt norm as
\begin{equation}\label{hs}
    \| G_k f(x) \|_2  = \| G_k f(x) \chi_m \|_2 + \lim_{N \to \infty } \Bigl\|  \sum_{n=m+1}^N G_k
    \chi_n (\chi_n - \chi_{n-1} )    f(x)  \Bigr\|_2
\end{equation}
Using (\ref{remnix}) and that $f \in L^2_{loc}$ we get that the first term on the rhs of (\ref{hs}) is bounded by a constant independent of $k$. Setting $K = \sup_{|x| \geq m} \left( |x|^{3+\varepsilon} |f(x)|\right)$, which is finite by condition of the lemma, for the second term we obtain
\begin{gather}
\lim_{N \to \infty } \Bigl\|  \sum_{n=m+1}^N G_k
    \chi_n (\chi_n - \chi_{n-1} )    f(x)  \Bigr\|_2  \leq K \sum_{n=m+1}^\infty \Bigl\|  G_k \chi_n (\chi_n - \chi_{n-1} )  \: |x|^{-3-\varepsilon} \Bigr\|_2 \\
    \leq K \sum_{n=m+1}^\infty \Bigl\|  G_k \chi_n \Bigr\|_2 (n-1)^{-3-\varepsilon} \leq  K b \sum_{n=m+1}^\infty n^2 (n-1)^{-3-\varepsilon} \label{rhs} ,
\end{gather}
where we have used (\ref{a1a}). The series on the rhs of (\ref{rhs}) obviously converge. Thus both terms on the rhs in
(\ref{hs}) are bounded by constants independent of $k$ and the lemma is proved.  \end{proof}

\section{Borromean Systems and the Multi-Particle Decay.}\label{sec:sec4}

In this section we shall deal with the systems, where for all values of $Z$  the bottom of the continuous spectrum is determined by the decay into $N$ single particles, which, in turn, means that $E_{thr} (Z) = 0$. Let us start with a couple of definitions. The system
of particles described by the Hamiltonian $H$ given in (\ref{xc31})--(\ref{:xc31}) (Z is fixed) is called {\em
Borromean} \cite{zhukov,fedorov} if neither of its subsystem has a bound state with the energy less or equal to zero.
(It is clear from the HVZ theorem \cite{reed} that the lowest
dissociation threshold of such system is given by the decay into $N$
single particles).

Let us introduce the Hamiltonians $H_{s}$ for $s = 1,2,\ldots, N$
\begin{equation}\label{sub}
    H_{s} = H_0 + \sum_{\substack{i<k \\ i,k \neq s}} V_{ik},
\end{equation}
which is the Hamiltonian of the original system, except that all
interactions of particle $s$ with other particles are set to zero.
We shall call the Hamiltonian $H$ \emph{strictly Borromean with a bound $\varepsilon$} if
\begin{equation}\label{crcoup}
    \varepsilon := \sup_p \bigl\{ p \in \mathbb{R}|\: \forall s \colon H_{s} \geq p \sum_{\substack{i<k \\ i,k \neq s}} \left( V_{ik}
    \right)_- \bigr\} >0
\end{equation}
The inequality (\ref{crcoup}) means that increasing the negative parts of
pair potentials in $H$ by small portions does not produce bound states in any of the
subsystems. Clearly, if $H$ is strictly Borromean then it is also Borromean but not vice versa.

\begin{theorem}\label{striborro}
Suppose that $(H(Z), \mathcal{Z})$ satisfies R1-4 and for all $Z \in \mathcal{Z}$ the Hamiltonian $H(Z)$ is strictly Borromean with a bound larger than some fixed $\varepsilon_0 >0$. Suppose, additionally, that all pair potentials satisfy the following inequality
\begin{equation}\label{many}
 2 \mu_{ik} V_{ik} (Z) \geq \frac {3+\delta}{4|x_i - x_k |^2} \quad
 \mathrm{if} \quad |x_i - x_k | \geq R_0 ,
\end{equation}
where $\delta > 0, R_0 \geq 0 $ are fixed constants independent of
$Z$ and $\mu_{ik}$ denotes the reduced mass of the particle pair
(i,k). Then: (a) for $Z_k \to Z_{cr}$ the sequence $\psi(Z_k)$ defined by R3 does
not spread. (b) $H(Z_{cr})$ has at least one bound state
with the zero energy; (c) $H(Z)$ has a finite number of bound states
with the energies less or equal to zero for all $Z \in \mathcal{Z}$.
\end{theorem}

\underline{\emph{Remark1:}} In the appendix we show that after strengthening the requirement R2 the condition in Theorem~\ref{striborro} for $H(Z)$ being strictly Borromean with a bound larger than $\varepsilon_0$ becomes too restrictive. It suffices that $H(Z)$ is Borromean for all $Z \in \mathcal{Z}$.

\underline{\emph{Remark2:}} Theorem~\ref{striborro} and its proof below hold also if some particles are bosons or fermions. (In this case $E_{thr} (Z)$ in R3 is defined as the bottom of the continuous spectrum of $\pi H \pi$, where $\pi$ defines the projection operator on the space with a proper permutation symmetry).

\begin{proof}[Proof of Theorem~\ref{striborro}]
Theorem~\ref{th:1} provides the inclusion $(a) \Rightarrow (b)$. Let us prove (a). By Lemma~\ref{rsd5} it is sufficient to show that every weakly converging subsequence of $\psi(Z_k)$ converges in norm. So let $\{Z_n\}_{n=1}^\infty \subset \mathcal{Z}/Z_{cr}$, $Z_n \to Z_{cr}$ be a subsequence such that $\psi_n := \psi(Z_n )$ is weakly convergent: $\psi_n \wwto \phi_0$, where $\phi_0 \in D(H_0)$ by Lemma~\ref{kuksi}. Our
aim is to obtain an upper bound on $|\psi_n|$ in the form
required by Theorem~\ref{rsd2}. This would mean that $|\psi_n|$ does not spread and hence by Lemma~\ref{rsd3} $\psi_n \to \phi_0$ in norm as needed.

In the following we derive the upper bound on the weakly converging
subsequence $\psi_n \wwto \phi_0$. For economy of notation we shall denote
$V^n_{ik} := V_{ik} (Z_n)$ and $E(Z_n )=-k_n^2 $ (where $k_n > 0$ and $k_n \to 0$ for $n\to \infty$). Let us start by writing out the Schr\"odinger equation
\begin{equation}\label{schro}
    \Bigl[ H_0 + k_n^2 + \sum_{i<k} \left( V^n_{ik}  \right)_+ \Bigr]
    \psi_n
    = \sum_{s<l} \left( V^n_{sl} \right)_- \psi_n
\end{equation}
Taking the inverse of the operator on the left we rewrite it
equivalently as
\begin{gather}
    \psi_n = \sum_{s<l} \Bigl[ H_0 + \sum_{i<k} \left( V^n_{ik}
\right)_+ + k^2_n \Bigr]^{-1} \left( V^n_{sl} \right)_- \psi_n = \label{schro2}\\
\sum_{s<l}  \Bigl[ H_0 + \sum_{i<k} \left( V^n_{ik} \right)_+ +
k^2_n \Bigr]^{-1} \left( V^n_{sl} \right)_-  (\psi_n - \phi_0) +
\Bigl[ H_0 + \sum_{i<k} \left( V^n_{ik} \right)_+ + k^2_n
\Bigr]^{-1}
      \left( V^n_{sl} \right)_- \phi_0 \label{schro2'}
\end{gather}
From (\ref{many}) the following inequality holds for the positive
parts of the pair potentials
\begin{equation}\label{many2}
( V^n_{ik} )_+ \geq \frac {3+\delta}{8 \mu_{ik} |x_i - x_k |^2}
\chi_{\{x|\: |x_i - x_k | \geq R_0 \}} .
\end{equation}
Let us introduce the resolvent
\begin{equation}\label{resol}
\mathcal{R} (z ) =     \left[ H_0 + \left( \sum_{i<k} \frac
{3+\delta}{8 \mu_{ik} |x_i - x_k |^2} \chi_{\{x|\: |x_i - x_k |
\geq R_0 \}} \right)- z \right]^{-1} ,
\end{equation}
which is a positivity preserving operator~ \cite{reed}. Using (\ref{many2}) and applying Lemma~\ref{bond} we obtain the following inequality out
of (\ref{schro2}-\ref{schro2'})
\begin{equation}
    |\psi_n | \leq  \sum_{s<l} \mathcal{R} (-k_n^2 )
      \left( V^n_{sl} \right)_- \bigl| \psi_n - \phi_0\bigr| + \sum_{s<l}
      \mathcal{R} (-k_n^2 )\left( V^n_{sl} \right)_- \bigl| \phi_0\bigr|
\end{equation}
The last inequality gives us the upper bound on $|\psi_n|$ in the
form
\begin{equation}\label{missed}
    |\psi_n| \leq g_n + |h_n| ,
\end{equation}
where
\begin{gather}
    g_n := \sum_{s<l} \mathcal{R} (-k_n^2 )
      \left( V^n_{sl} \right)_- \bigl| (\psi_n - \phi_0) \bigr| \label{.1}\\
      h_n := \sum_{s<l} \mathcal{R} (-k_n^2 )\chi_{\{x|\; |x_s -x_l| \leq R_0\}}
      F_{sl}  |\phi_0| \label{.2},
\end{gather}
where in (\ref{missed}) we have used the inequality
\begin{equation}\label{ins}
\left( V^n_{sl} \right)_-  \leq \chi_{\{x|\; |x_s -x_l| \leq R_0\}}
F_{sl}
\end{equation}
It remains to prove that $g_n, h_n$ defined by
(\ref{.1})--(\ref{.2}) indeed satisfy the requirements of
Theorem~\ref{rsd2}. Using (\ref{ins}) for $g_n$ we get
\begin{gather}\label{gn}
    \| g_n \| \leq  \sum_{s<l} \Bigl\| \mathcal{R} (-k_n^2 )\chi_{\{x|\; |x_s -x_l| \leq R_0\}}
      F^{1/2}(x_s - x_l ) \Bigr\| \: \bigl\| \left( V^n_{sl} \right)^{1/2}_- (\psi_n -
      \phi_0)\bigr\|
\end{gather}
Applying Lemmas~\ref{dos1}, \ref{dos2} to (\ref{gn}) we get that
$\|g_n \| \to 0$ for $n \to \infty$ as needed. The following inequality holds for the norm of
$h_n$
\begin{gather}\label{hn}
    \| h_n \| \leq  \sum_{s<l} \Bigl\| \mathcal{R} (-k_n^2 )\chi_{\{x|\; |x_s -x_l| \leq R_0\}}
      F^{1/2}(x_s - x_l ) \Bigr\| \: \bigl\| F^{1/2}(x_s - x_l )
      \phi_0\bigr\|
\end{gather}
Because $\phi_0 \in D(H_0 )$ the term $\bigl\| F^{1/2}(x_s - x_l )
\phi_0\bigr\|$ is finite. The operator norm on the rhs of (\ref{hn})
is uniformly bounded by Lemma~\ref{dos2}. So, on one hand, $\| h_n
\| $ is uniformly bounded as needed. On the other hand, looking at
the definition of $h_n$ given by (\ref{.2}) and using
Lemma~\ref{bond} it is easy to see that $|h_n | \leq |h_{n'}|$ if
$k^2_n \geq k^2_{n'} $. Because $k^2_n > 0$ and at the same time $k^2_n \to 0$ from
any subsequence $k^2_{n_{s}}$ one can extract a monotonically
decreasing sub/subsequence. Hence, from any subsequence
$h_{n_{s}}$ one can extract a sub/subsequence, which is
non--decreasing. Thus both $h_n$ and $g_n$ satisfy the
requirements of Theorem~\ref{rsd2} and (a) and (b) are proved.

Let us prove (c). A proof by contradiction. Suppose that for some
fixed $Z$ there is an infinite number of orthonormal bound
states $\phi_n$ with the energy $E_n = -k^2_n$ less or equal to
zero. Let us first assume that $E_n <0$ (all energies are
negative). The energies must accumulate at zero, so $k^2_n \to 0$.
Because the sequence $\phi_n$ is orthonormal $\phi_n \wwto
0$. Just repeating the arguments above we shall get the upper
bound
\begin{equation}
|\phi_n | \leq \sum_{s<l} \mathcal{R} (-k_n^2 )
      \left( V_{sl} \right)_- \bigl| \phi_n \bigr|
      \label{.12}
\end{equation}
It is easy to see that Lemma~\ref{dos1} applies in this case as well and gives
\begin{equation}\label{suc:scu}
\lim_{n \to \infty}\left\| \left( V_{sl} \right)^{1/2}_- \bigl| \phi_n \bigr| \right\| =  0
\end{equation}
(To formally apply Lemma~\ref{dos1} to this case one can redefine $Z_n = k^2_n$ and $Z_{cr} = 0$, which makes $H$ independent of $Z_n$). Applying Lemma~\ref{dos2} to (\ref{.12}) and using (\ref{suc:scu}) we infer that the norm of the rhs of (\ref{.12}) goes to zero. This means that $\|\phi_n\|
\to 0$, which is nonsense. Thus the number of bound states with
negative energies must be finite. It remains to show that the
number of bound states with the zero energy is finite. Again, we assume by contradiction that this number is infinite. Let us replace the pair potentials $V_{ik}$ through $\tilde V_{ik} = V_{ik} -
(\varepsilon_0/2)(V_{ik})_-$. On one hand, the system with the redefined potentials satisfies all conditions of the present theorem. Indeed, it is strictly Borromean with a bound larger than $\varepsilon_0/2$ and (\ref{many}) still holds because the inequality (\ref{many}) concerns only positive parts of pair potentials. On the other hand, this change in the potentials makes the number of bound states with negative energy infinite. But we have already shown that this number must be finite. This is a contradiction.  \end{proof}

The following simple lemma is similar to the one in
 \cite{we}.
\begin{lemma}\label{bond}
Let $V^{(1)}, V^{(2)}$ be the sums of pair interactions satisfying R1 and
suppose that $V^{(1)} \geq V^{(2)}$. Let $\mathcal{R}_{1,2} = \left[H_0 + V^{(1,2)} - z_{1,2}\right]^{-1}$ denote the resolvents with resolvent sets $\rho_{1,2}$. Then for $z_i  \in \mathbb{R} \cap \rho_i$ and $z_1 \leq z_2 $  one has the inequality
\begin{equation}\label{eqbond}
\mathcal{R}_1 |f| \leq \mathcal{R}_2 |f|    \quad  (\forall f \in L^2  )
\end{equation}
\end{lemma}
\begin{proof}
By the second resolvent formula
\begin{equation}\label{bond2}
        \mathcal{R}_2 |f| - \mathcal{R}_1 |f| = \mathcal{R}_1  \left[V^{(1)} -
        V^{(2)} + (z_2 - z_1)\right] \mathcal{R}_2 |f| \geq 0
\end{equation}
because the expression in the square brackets is non--negative and the operators
$\mathcal{R}_{1,2}$ are positivity preserving \cite{reed}.
 \end{proof}

\begin{lemma}\label{dos1} Suppose that $(H(Z), \mathcal{Z})$ satisfies R1-4 and for all $Z \in \mathcal{Z}$ the Hamiltonian $H(Z)$ is strictly Borromean with a bound larger than some fixed $\varepsilon_0 >0$. Suppose, additionally, that $\psi(Z_n)
\wwto \phi_0 \in D(H_0)$, where $Z_n \to Z_{cr}$ and $Z_n \in \mathcal{Z}/Z_{cr}$. Then $\| \left( V_{sl} (Z_n) \right)^{1/2}_-
(\psi(Z_n) - \phi_0 )  \| \to 0$.
\end{lemma}
\begin{proof}
By the IMS formula \cite{ims,teschl} (for a detailed derivation see \cite{teschl}) the
Hamiltonian $H(Z)$ can be decomposed in the following way
\begin{equation}\label{ims}
    H (Z) = \sum_{s=1}^N J_s H_s (Z) J_s + K(Z)
\end{equation}
where
\begin{gather}
K(Z) = \sum_{s=1}^N I_s (Z) |J_s |^2 + \sum_{s=1}^N  |\partial J_s |^2 \label{K} \\
I_s (Z) = \sum_{i=1}^N V_{is} (Z)
\end{gather}
The functions of the IMS decomposition satisfy $J_s \in C^{\infty}
(\mathbb{R}^{3N-3})$, $0 \leq J_s \leq 1$ and form the partition
of unity $\sum_s J^{2}_s =1$. Besides $J_s$ are homogeneous of
degree zero outside the unit sphere, {\em i.e.} $J_s (\lambda x) =
J_s (x)$ for $\lambda \geq 1$ and $|x|= 1$ (this makes $|\partial
J_s |$ fall off at infinity), and there exists $C > 0$ such that
\begin{equation}\label{ims5}
    \supp J_s \cap \{ x | |x| > 1 \} \subset \{x|\; |x_i - x_s | \geq C |x| \quad
    \text{for} \quad i \neq s \}
\end{equation}

For a shorter notation we denote $\psi_n := \psi(Z_n)$. We shall prove the lemma in three steps given by the following equations.
\begin{gather}
(a) \quad \lim_{Z_n \to Z_{cr}} \left((\psi_n - \phi_0 ) , K(Z_n) (\psi_n - \phi_0 ) \right) = 0 \label{kg0} \\
(b) \quad \lim_{Z_n \to Z_{cr}} \left((\psi_n - \phi_0 )  , H(Z_n)  (\psi_n - \phi_0 )  \right)= 0 \label{kg0:b}\\
(c) \quad \lim_{Z_n \to Z_{cr}} \left((\psi_n - \phi_0 )  , \left( V_{sl}
(Z_n) \right)_-  (\psi_n - \phi_0 )  \right)= 0 \label{kg0:c}
\end{gather}
From $(c)$ the statement of the lemma clearly follows. Let us
start with $(a)$. By the requirement R1 on pair potentials we have
\begin{equation}\label{tobe}
 |(f,K(Z) f)| \leq (f,\tilde K f) \quad \quad (\forall f \in D(H_0) )
\end{equation}
where the operator $\tilde K$ is defined through
\begin{gather}
\tilde K = \sum_{s=1}^N \tilde I_s |J_s |^2 + \sum_{s=1}^N  |\nabla J_s |^2 \label{tilk1} \\
\tilde I_s = \sum_{i=1}^N F_{is}  \label{tilk2}
\end{gather}
Eq.~(\ref{tobe}) has the advantage that $\tilde K$ does not depend
on $Z$ and to prove $(a)$ it suffices to show that
\begin{equation}\label{j233}
((\psi_n - \phi_0 ) , \tilde K (\psi_n - \phi_0 ) ) \to 0
\end{equation}
Eq.~(\ref{j233}), in turn, would follow from Lemma~\ref{kuksi} if we would prove that $\tilde K$ is relatively $H_0$ compact. Because $ |\nabla J_s |^2 \in L^\infty_\infty (\mathbb{R}^{3N-3})$ the second sum in (\ref{tilk1}) represents an operator, which is, indeed, relatively $H_0$ compact (see Lemma 7.11
in \cite{teschl}). It remains to demonstrate that the operators $\tilde I_s |J_s |^2 $ are also relatively
$H_0$ compact. Indeed, we can write
\begin{equation}\label{tobe2}
F_{is} |J_s |^2   =  F_{is} |J_s |^2 \chi_{\{|x| \geq
n\}} + F_{is} |J_s |^2 \chi_{\{x| \; |x| < n\}}
\end{equation}
By (\ref{ims5}) we can fix $n$ in
(\ref{tobe2}) so that $\supp \left[F_{is} |J_s |^2
\chi_{\{|x| \geq n\}} \right] \subset \{x|\; |x_i -x_s | \geq
d_0\}$. Because $F_{is} \chi_{\{x|\; |x_i -x_s | \geq d_0\}}
\in L^\infty_\infty$ this would mean that the first term on the
rhs of (\ref{tobe2}) is relatively $H_0$ compact. The second term belongs $L^2
(\mathbb{R}^{3N-3})$ and thus is also relatively $H_0$ compact.

This makes us conclude that $\tilde K$ is relatively $H_0$ compact
as a sum of relatively $H_0$ compact operators and (\ref{j233}) holds by
Lemma~\ref{kuksi}. This proves (a).

Let us prove $(b)$. Rewriting the expression in (b) we obtain
\begin{gather}
\bigl( (\psi_n - \phi_0 ), H(Z_n) \: (\psi_n - \phi_0 )\bigr)
= E(Z_n) \bigl((\psi_n - \phi_0 ), \psi_n\bigr) - \label{tobe342}\\
\bigl((\psi_n -\phi_0 ), H(Z_{cr})\phi_0 \bigr)  - \bigl((\psi_n -\phi_0 ), \left[V(Z_n)-V(Z_{cr})\right]\phi_0 \bigr) \label{tobe34}
\end{gather}
where we have used the equations $H(Z_n) \psi_n = E(Z_n)\psi_n$ and $H(Z_n) = H(Z_{cr})
+ V(Z_n) - V(Z_{cr})$. The first term on the rhs of (\ref{tobe342})--(\ref{tobe34}) goes to zero because $E(Z_n) \to 0$. The second term goes to zero because $H(Z_{cr})\phi_0 $ is a fixed function in $L^2$ and $\psi_n \wwto \phi_0$. The statement (b) would be proved if we would show that the third term goes to zero. Because $C_0^\infty (\mathbb{R}^{3N-3})$ is dense in $D(H_0)$ we can decompose $\phi_0$ in the following way $\phi_0 = \phi_0' + \phi_0''$, where $\phi_0' \in C_0^\infty (\mathbb{R}^{3N-3})$ and $\| \phi_0'' \|$ can be made as small as pleased. Then the third term on the rhs of (\ref{tobe342})--(\ref{tobe34}) takes the form
\begin{gather}
\bigl((\psi_n -\phi_0 ), \left[V(Z_n)-V(Z_{cr})\right]\phi_0 \bigr) = \bigl((\psi_n -\phi_0 ), \left[V(Z_n)-V(Z_{cr})\right]\phi_0' \bigr) +  \label{srd}\\
\bigl(\left[V(Z_n)-V(Z_{cr})\right](\psi_n -\phi_0 ), \phi_0'' \bigr) \label{srd2}
\end{gather}
The first term on the rhs of (\ref{srd})--(\ref{srd2}) goes to zero by R2. The second term could be made as small as pleased if we would show that $\left[V(Z_n)-V(Z_{cr})\right](\psi_n -\phi_0 )$ is a uniformly bounded sequence. But this can be deduced from the following inequality
\begin{equation}
\Bigl\| \left[V(Z_n)-V(Z_{cr})\right](\psi_n -\phi_0 )\Bigr\| \leq a \| H_0 (\psi_n - \phi_0) \| + b \label{sfd},
\end{equation}
where the constants $a,b$ are independent of $Z$ (cf. Eqs.~(\ref{kuka})--(\ref{kuka:1})). The lhs of (\ref{sfd}) is uniformly bounded because $(\psi_n - \phi_0)$ is an uniformly $H_0$-bounded sequence by Lemma~\ref{forg}.

It remains to be shown that $(c)$ is true. Using the statement (a) and (\ref{ims}) we obtain from $(b)$
\begin{equation}\label{tobe4}
\bigl((\psi_n - \phi_0 ), J_s  H_{s} (Z_n) J_s (\psi_n - \phi_0 )  \bigr)
\to 0 \quad \quad (\forall s)
\end{equation}
Together with the inequality (\ref{crcoup}) for a strictly
Borromean system this gives us
\begin{equation}\label{kotz1}
    \bigl((\psi_n - \phi_0 )  , J^{2}_s \bigl(V_{ik} (Z_n) \bigr)_- (\psi_n - \phi_0 ) \bigr) \to
    0 \quad \quad (\forall s, \forall i,k \neq s)
\end{equation}
Finally, using (\ref{kotz1}) we obtain
\begin{gather}
    \bigl((\psi_n - \phi_0 )  , \bigl(V_{ik} (Z_n)\bigr)_- (\psi_n - \phi_0 ) \bigr) = \sum_{s=1}^N  \bigl((\psi_n -
    \phi_0 )  , J^{2}_s
    \bigl(V_{ik} (Z_n)\bigr)_- (\psi_n - \phi_0 ) \bigr) \label{kotz112}\\
\to     \bigl((\psi_n - \phi_0 )  , (J^{2}_i + J^{2}_k
)\bigl(V_{ik} (Z_n)\bigr)_- (\psi_n - \phi_0 )    \bigr)
\label{kotz2}
\end{gather}
On the other hand, for the rhs of (\ref{kotz2}) we get
\begin{gather}
    0 \leq \bigl( (\psi_n - \phi_0 )  , ( J^{2}_i + J^{2}_k )  \bigl( V_{ik} (Z_n)\bigr)_- (\psi_n - \phi_0 ) \bigr) \label{hand1}\\
    \leq \bigl((\psi_n - \phi_0 )  , ( J^{2}_i + J^{2}_k ) F_{ik} (\psi_n -
\phi_0 ) \bigr) \to 0 \label{hand2}
\end{gather}
because $( J^{2}_i + J^{2}_k ) F_{ik} $ is relatively $H_0$
compact and Lemma~\ref{kuksi} applies. Eqs.
(\ref{hand1})--(\ref{hand2}) mean that the rhs respectively lhs of
(\ref{kotz112})--(\ref{kotz2}) go to zero and thus (c) is
proved. \end{proof}

\begin{lemma}\label{dos2}
The following inequality holds for the resolvent defined in
(\ref{resol})
\begin{equation}\label{bibo}
\sup_{k>0} \Bigl\| \mathcal{R} (-k^2) F_{sl}^{1/2}
\chi_{\{x| \: |x_s - x_l| \leq n \}}\Bigr\| < \infty \quad \quad
(\forall s\neq l, \forall n>0)
\end{equation}
\end{lemma}
\begin{proof}
Without loosing generality let us set $s = 1$ and $l =2$. It is
convenient \cite{greiner} to choose the Jacobi set of orthogonal
coordinates $\xi = x_1 - x_2 $ and $\zeta = (\zeta_1 , \ldots ,
\zeta_{N-2})$, where $\xi \in \mathbb{R}^{3}$ and $\zeta \in
\mathbb{R}^{3N-6}$ and consider the problem on $L^2
(\mathbb{R}^{3N-3}) = L^2 (\mathbb{R}^{3}) \otimes L^2
(\mathbb{R}^{3N-6})$. It is always possible to choose the
$\zeta$--coordinates so that the kinetic energy operator takes the
form
\begin{equation}\label{kinen}
    H_0 = \frac{p_{\xi}^2}{2 \mu_{12}}  - \Delta_{\zeta},
\end{equation}
where $p_{\xi}$ is the conjugate momentum for $\xi$ and $\Delta_{\zeta}$ denotes the Laplacian on the space of ${\zeta}$ coordinates.

For a shorter notation let us denote
\begin{gather}
\tilde F_{sl} := F_{sl}^{1/2} \chi_{\{x| \: |x_s - x_l|
\leq n \}} \\
\eta_{sl} := \frac{3+\delta}{4 |x_s - x_l |^2} \chi_{\{x|\:
|x_s - x_l | \geq R_0 \}}
\end{gather}
Let us choose arbitrary $f \in L^2 $ with $\|f\|=1$. Consecutively
applying Lemma~\ref{bond} one obtains the following chain of
inequalities
\begin{gather}
\left\| \mathcal{R} (-k^2)
      \tilde F_{12}^{1/2}  f \right\| \leq \left\| \mathcal{R} (-k^2)
      \tilde F_{12}^{1/2}  |f| \right\|
      \leq \left\| \Bigl[ H_0 + k^2 +  (2\mu_{12})^{-1} \eta_{12} \Bigr]^{-1}
      \tilde F_{12}^{1/2}  |f|\right\| \label{abs} \\
            =  \left\| U \Bigl[ H_0 + k^2 + (2\mu_{12})^{-1} \eta_{12}
    \Bigr]^{-1}
      \tilde F_{12}^{1/2}  |f|\right\| \label{abs2}
\end{gather}
where the unitary operator $U = 1 \otimes \mathcal{F}$ has been
introduced, $\mathcal{F}$ being the Fourier transform in the space
of $\zeta$--coordinates $\mathcal{F} g (\zeta) = \hat g (\lambda
)$. Using that $U$ commutes with $\tilde F_{12}^{1/2}$ and
$\eta_{12}$ we can continue the right--hand side of (\ref{abs2})
as
\begin{gather}
= \left\| \Bigl[ (2\mu_{12})^{-1} p_{\xi}^2  + \lambda^2 + k^2 +
(2\mu_{12})^{-1} \eta_{12} \Bigr]^{-1}
      \tilde F_{12}^{1/2} U |f|\right\| \\ \leq
      2\mu_{12} \left\| \Bigl[    p_{\xi}^2  +
      \eta_{12}   +  2\mu_{12} k^2    \Bigr]^{-1}
      \tilde F_{12}^{1/2}   \otimes 1 \: \bigl| U
      |f| \bigr| \: \right\| \label{line}
\end{gather}
where the first term of the tensor product is an operator acting
on $\xi$-space, which is $L^2 (\mathbb{R}^3 )$. Because  $\left\|
\bigl| U |f| \, \bigr| \right\| = 1$ we finally get from
(\ref{abs})--(\ref{line})
\begin{equation}\label{ugme}
\left\| \mathcal{R} (-k^2)
      \tilde F_{12}^{1/2}  \right\| \leq
      2\mu_{12} \left\| \Bigl[    p_{\xi}^2  +
      \eta_{12}   +  2\mu_{12} k^2    \Bigr]^{-1}
      F^{1/2} (\xi) \chi_{\{\xi| \: |\xi| \leq n \}}\right\|,
\end{equation}
where the norm on the rhs is that of the operator acting on $L^2
(\mathbb{R}^3 )$. In Sec.~\ref{sec:sec3} we have shown that the
maximum of the rhs of (\ref{ugme}) over $k>0$ is finite. This
proves the lemma.
 \end{proof}

\section{The Two-Cluster Decay}\label{sec:sec5}

Here we consider the situation, where the bottom of the continuous spectrum corresponds to the dissociation into two clusters. Let $a = 1, 2,
\ldots, (2^{N-1}-1)$ label all the distinct ways \cite{ims} of partitioning
particles into two non--empty clusters $\mathfrak{C}^{a}_1$ and $\mathfrak{C}^{a}_2$. It is useful to construct the following operators $H_a = H_0 + V_{1}^a + V_{2}^a$, where $V_{1,2}^a$ denote the sum of
interactions in the corresponding cluster
$\mathfrak{C}^{a}_{1,2}$ (in other words, $V_{1}^a + V_{2}^a$ is the sum of all
interactions minus cross-terms between the two clusters).

We shall assume that
for all $Z \in \mathcal{Z}$ the lowest dissociation threshold is formed by the decay into two particular
clusters $\mathfrak{C}^1_1$ and $\mathfrak{C}^1_2$, which correspond
to the partition $a = 1$.
Let $\xi \in \mathbb{R}^{3N - 6}$ denote the
set of internal coordinates in both clusters $\mathfrak{C}^1_{1,2}$ and $r \in
\mathbb{R}^3$ is the coordinate of clusters' relative motion ($r$
points from the center of mass of $\mathfrak{C}_1$ to the center
of mass of $\mathfrak{C}_2$).

The Hamiltonian of the system given by (\ref{xc31})--(\ref{:xc31})
can be rewritten as
\begin{equation}\label{goham}
    H = H_{thr} (\xi, Z) + \frac{p^{2}_r }{2 \mu } + W (r, \xi, Z)
\end{equation}
where $H_{thr}$ is the Hamiltonian of internal motion in the clusters
$\mathfrak{C}^1_{1,2}$, $p_r$ is the conjugate momentum
corresponding to the coordinate $r$, $\mu = M_1 M_2 / (M_1 + M_2
)$ is the reduced mass derived from clusters' total masses $M_{1,2}$ and
$W$ is the sum of pair interactions between the clusters (cross--terms). It is convenient to consider (\ref{goham}) on a tensor product space $L^2 (\mathbb{R}^{3N-3}) = L^2 (\mathbb{R}^{3}) \otimes L^2 (\mathbb{R}^{3N-6})$, where each term in the tensor product corresponds to the space of $r$ and $\xi$ coordinates respectively.

Apart from requirements R1-4 on the pair potentials the
following additional requirements have to be imposed
\begin{list}{R\arabic{foo}}
{\usecounter{foo}
    \setlength{\rightmargin}{\leftmargin}}
\setcounter{foo}{4}
\item
For all $Z\in \mathcal{Z}$ there exist a normalized bound state $\phi_{thr} (\xi, Z) \in D(H_{thr})$ and a constant $|\Delta \epsilon| > 0$ independent of $Z$ such that $H_{thr} (Z) \phi_{thr} (\xi, Z) = E_{thr}(Z) \phi_{thr} (\xi, Z)$ and
\begin{gather}
\bigl(1-P_{thr}(Z)\bigr) \Bigl[H_{thr}(Z) - E_{thr}(Z)\Bigr]
\geq |\Delta \epsilon| \bigl(1-P_{thr}(Z)\bigr) \label{need1}\\
\Bigl[ H_{a\geq 2} (Z) - E_{thr}(Z)\Bigr] \geq |\Delta \epsilon|,
\label{need2}
\end{gather}
where $P_{thr} (Z) = 1 \otimes \phi_{thr}(\phi_{thr}, \cdot )$ is a projection operator.
\item
For all $Z \in \mathcal{Z}$ there are $A,q > 0$ independent of $Z$ such that the bound state $\phi_{thr}$ defined in $R5$ satisfies the following inequality $|\phi_{thr} (\xi, Z) | \leq A e^{-q|\xi|}$.
\end{list}

The following remarks are due. The requirement R5 says that the bottom of the continuous spectrum corresponds to the ground state of $H_{thr}$. This ground state is non-degenerate \cite{reed} and consequently $\phi_{thr} (\xi, Z) \geq 0$. The requirement R6 is necessary to control the exponential fall off of the clusters' ground state. Examples of  upper bounds on the wave function giving the control of this kind can be found in Refs.~ \cite{osten}.

From the requirement R6 we infer the following inequality
\begin{equation}\label{pexpi}
\bigl| P_{thr} (Z) f \bigr| \leq P_{thr}(Z) |f| \leq P_{exp} |f|  \quad
\quad (\forall f \in L^2),
\end{equation}
where $P_{exp} = 1\otimes A^2 e^{-q|\xi|} (e^{-q|\xi|} , \cdot )$ (the operator $P_{exp}$ is not a projection operator and is introduced merely for convenience).

\begin{theorem}\label{th2}
Suppose that $(H(Z), \mathcal{Z})$ satisfies R1-6 and for all $Z \in \mathcal{Z}$ the potentials satisfy the following inequality
\begin{equation}\label{W}
    2 \mu W \geq \frac{3 + \delta}{4 |r|^2 } \quad {\rm if} \quad
    |r| \geq C_0 + C_1 |\xi|^p
\end{equation}
where $\delta , C_{0,1} , p$ are fixed positive constants. Then: (a) for $Z_k \to Z_{cr}$ the sequence $\psi(Z_k)$ defined by R3 does
not spread. (b) $H(Z_{cr})$ has at least one bound state at the bottom of the continuous spectrum.
\end{theorem}

\begin{proof}
Again, following the arguments of
Theorem~\ref{striborro}, the theorem would be proved if we would show that every weakly converging subsequence of the sequence $\psi (Z_k)$ does not spread. So let $\{Z_n\}_{n=1}^\infty \subset \mathcal{Z}/Z_{cr}$, $Z_n \to Z_{cr}$ be a subsequence such that $\psi_n := \psi(Z_n )$ is weakly convergent: $\psi_n \wwto \phi_0$, where $\phi_0 \in D(H_0)$ by Lemma~\ref{kuksi}. Our aim is to show that $\psi_n$ does not spread.

Using the obvious identity
\begin{equation}\label{obvi}
\psi_n = P_{thr} (Z_n) (\psi_n -\phi_0) + (1-P_{thr} (Z_n) )
(\psi_n -\phi_0) + \phi_0
\end{equation}
and the inequality (\ref{pexpi}) we obtain the upper bound on $|\psi_n|$
\begin{equation}\label{sex}
|\psi_n| \leq \bigl| (1-P_{thr} (Z_n) ) (\psi_n -\phi_0) \bigr| +
P_{thr} (Z_n) |\psi_n| + P_{exp} | \phi_0 | +  |\phi_0|.
\end{equation}
The last two terms on the rhs of
(\ref{sex}) are fixed $L^2$ functions and the first term goes to
zero in norm by Lemma~\ref{6}. Hence, to prove that $\psi_n$ does
not spread it suffices to prove that $P_{thr} (Z_n) |\psi_n|$ does
not spread. This is what we prove below.

Let us introduce the potential tail $\eta (x) = (3 + \delta )
(2r)^{-2} \chi_{\{x|\:|r| \geq 1\}}$.  We separate the potential
into two parts $W(Z_n) = W_n^+ - W_n^-$, where $W_n^+ := \max [ W(Z_n), (2
\mu)^{-1} \eta (r) ]$ and $W_n^- := W_n^+ - W(Z_n)$. Note that though
$W_n^+ \geq 0$, $W_n^-$ can take both signs, so $W_n^{\pm} $
are not positive and negative parts of the potential. To avoid
possible confusion, we distinguish the present notation by setting
the plus and minus signs as superscripts. If we abbreviate
$\Omega = \{ x | \: \abs{r} < C_0 + C_1 |\xi|^p \} $, then
according to (\ref{W}) $\text{supp} (W_n^- ) \subset \Omega$.

Rewriting the Schr\"odinger equation for $\psi_n$ we get
\begin{equation}\label{schr3}
    \psi_n = \Bigl[ (H_{thr} - E_{thr}) + \frac{p_{r}^2 }{2\mu }+  \frac{k_n^2}{2\mu} + W_n^+ \Bigr]^{-1} W_n^-
    \psi_n,
\end{equation}
where $k_n^2 = - 2\mu E(Z_n)$. Applying Lemma~\ref{bond} to (\ref{schr3}) gives
\begin{gather}
   \abs{\psi_n}
    \leq 2\mu \Bigl[ 2\mu (H_{thr} - E_{thr}) + p_{r}^2 + k_n^2 + \eta \Bigr]^{-1} \abs{W_n^-
    \psi_n}, \label{:avant}
\end{gather}
where we have used $2 \mu W_n^+ \geq \eta $. The advantage of
(\ref{:avant}) is that $P_{thr} $ commutes with $\eta$ and $p_r$ and besides $P_{thr}(H_{thr} - E_{thr})= 0$.
Thus, acting with $P_{thr} (Z_n)$ on both sides of (\ref{:avant}) gives us
\begin{gather}
P_{thr} (Z_n)   \abs{\psi_n} \leq 2\mu \Bigl[p_{r}^2 + k_n^2 + \eta
\Bigr]^{-1} P_{thr} (Z_n)\abs{W_n^-
    \psi_n} \label{avant1}\\
\leq 2\mu \Bigl[p_{r}^2 + k_n^2 + \eta \Bigr]^{-1} P_{exp} \:
\chi_{\{ \Omega \} } \abs{W_n^- \psi_n} \label{avant2}
\end{gather}
where we have used that $\text{supp} (W_n^- ) \subset \Omega$. Let
us consider the term $\abs{W_n^- \psi_n} $
\begin{equation}\label{pro}
\abs{W_n^- \psi_n} \leq \bigl| W_n^- (\psi_n - \phi_0 ) \bigr| +
\bigl| W_n^- \phi_0  \bigr| \leq \bigl| W_n^- (\psi_n - \phi_0 )
\bigr| + \left[\eta + 2 \sum_{i<j} F_{ij}\right] |\phi_0|,
\end{equation}
where we have used R1 together with the inequality $|W_n^- | \leq 2|W(Z_n)| + \eta $.
Substituting (\ref{pro}) into (\ref{avant2}) we finally obtain the
inequality
\begin{equation}
P_{thr} (Z_n)   \abs{\psi_n} \leq g_n + |h_n| ,
\end{equation}
where
\begin{gather}
g_n := 2\mu \Bigl[p_{r}^2 + k_n^2 + \eta \Bigr]^{-1} P_{exp} \:
\chi_{\{ \Omega \} } \bigl| W_n^-
(\psi_n - \phi_0 ) \bigr| \label{gn2} \\
h_n := 2\mu \Bigl[p_{r}^2 + k_n^2 + \eta \Bigr]^{-1} P_{exp}
\: \chi_{\{ \Omega \} }
|\tilde \phi_0| \label{hn2}\\
|\tilde \phi_0| := \left[\eta + 2 \sum_{i<j} F_{ij}\right]
|\phi_0| \in L^2
\end{gather}
To prove the theorem it remains to show that $g_n$ and $h_n$
defined by (\ref{gn2})--(\ref{hn2}) satisfy the conditions of
Theorem~\ref{rsd2}. Indeed, $\|g_n\| \to 0$ by the Lemmas~\ref{6},
\ref{7}. Let us consider $|h_n|$. Lemma~\ref{7} tells us
that the sequence $|h_n|$ is uniformly norm bounded (because
$|\tilde \phi_0| $ is a fixed $L^2$-function). By repeating the
arguments in the proof of Theorem~\ref{striborro} we find that from
any subsequence $|h_{n_{s}}|$ one can extract a non--decreasing sub/subsequence. Thus Theorem~\ref{rsd2} applies and
$P_{thr} (Z_n) \abs{\psi_n} $ does not spread.  \end{proof}

The following two lemmas supplement the proof of Theorem~\ref{th2}.

\begin{lemma}\label{6}
Suppose that $(H(Z), \mathcal{Z})$ satisfies R1-6 and $\psi(Z_n)
\wwto \phi_0 \in D(H_0)$, where $Z_n \to Z_{cr}$ and $Z_n \in \mathcal{Z}/Z_{cr}$. Then
\begin{gather}
(a) \quad \quad \lim_{Z_n \to Z_{cr}} \bigl\| W_n^- (\psi_n - \phi_0 )\bigr\| = 0 \\
(b) \quad \quad \lim_{Z_n \to Z_{cr}} \Bigl\| \bigl[1 - P_{thr}(Z_n)\bigr] (\psi_n - \phi_0)\Bigr\| = 0,
\end{gather}
where $\psi_n := \psi(Z_n)$ and $W_n^-$ is defined as in Theorem~\ref{th2}.
\end{lemma}

\begin{proof}
The IMS formula \cite{ims} reads
\begin{gather}
    H(Z) = \sum_{a=1}^{2^N -1} J_a H_a (Z) J_a + K(Z) \label{ims3} \\
    K(Z) = \sum_{a=1}^{2^N -1} \left[ J^{2}_a I_a(Z) + |\nabla J_a|^2 \right] ,
\end{gather}
where
\begin{equation}
I_a (Z) := \sum_{\substack{i \in \mathfrak{C}^{a}_1 \\j \in
\mathfrak{C}^{a}_2}} V_{ij} (Z)
\end{equation}
and the functions $J_a$ are defined exactly as in
 \cite{ims}.

Let us prove (a).
The first step is to prove the following
equations
\begin{gather}
\lim_{Z_n \to Z_{cr}} \bigl(\psi_n - \phi_0 , K(Z_n ) (\psi_n -
\phi_0 )\bigr) = 0
\label{kpsi}\\
\lim_{Z_n \to Z_{cr}}  \bigl\| J_{a \geq 2} (\psi_n - \phi_0 )
\bigr\| = 0 \label{Ja}
\end{gather}
Similarly to the proof of Theorem~\ref{striborro}, we introduce the
operator
\begin{equation}\label{tildek}
    \tilde K = \sum_a \left[ J^{2}_a
    \sum_{\substack{i \in \mathfrak{C}^{a}_1 \\j \in \mathfrak{C}^{a}_2}}F_{ij} + |\nabla J_a|^2
    \right],
\end{equation}
which does not depend on $Z$ and is relatively $H_0$ compact.
Using the inequality (\ref{tobe}), which is also true for the
redefined $K, \tilde K$ and applying Lemma~\ref{kuksi} we find
that (\ref{kpsi}) holds.

Just repeating the argumentation in Lemma~\ref{dos1} (around
Eq.~(\ref{tobe34})) we obtain
\begin{equation}\label{zuzu}
\lim_{Z_n \to Z_{cr}} \bigl( (\psi_n-\phi_0), \left[H(Z_n) -
E_{thr}(Z_n)\right] (\psi_n-\phi_0) \bigr) = 0.
\end{equation}
Substituting (\ref{ims3}) into (\ref{zuzu}) and using (\ref{kpsi})
we obtain
\begin{equation}\label{proob}
\lim_{Z_n \to Z_{cr}} \sum_{a} \bigl( J_a (\psi_n - \phi_0),
\left[H_a (Z_n) - E_{thr}(Z_n)\right] J_a (\psi_n - \phi_0)\bigr) = 0
\end{equation}
When $a=1$ the following inequality holds
\begin{equation}\label{neznaja}
   H_1 (Z_n) = \frac{p_r^2}{2\mu} + H_{thr}(Z_n) \geq E_{thr}(Z_n)
\end{equation}
Setting (\ref{need2}) into (\ref{proob}) and using (\ref{neznaja}) we obtain (\ref{Ja}).

Because $|W_n^- | \leq 2|W| + \eta $ to prove (a) it is enough to
show that
\begin{align}
 &\bigl\|F_{ij} (\psi_n - \phi_0) \bigr\| \xrightarrow[n \to \infty]{} 0  \quad
\quad
(i \in \mathfrak{C}^1_1 , \: \: j \in \mathfrak{C}^1_2) \label{rema}\\
 &\| \eta (\psi_n - \phi_0) \| \xrightarrow[n \to \infty]{} 0 \label{rema2}
\end{align}
For the last expression we have
\begin{equation}\label{nach1}
\| \eta (\psi_n - \phi_0) \|^2   = \| J_1 \eta (\psi_n - \phi_0)
\|^2 + \sum_{a\geq 2} \| J_a \eta (\psi_n - \phi_0) \|^2
\end{equation}
The first term on the rhs of (\ref{nach1}) goes to
zero because $J_1 \eta $ is bounded and falls off at infinity and is thus
relatively $H_0$ compact, and all other terms go to zero due to
(a) and $\eta $ being bounded. Hence, indeed (\ref{rema2})
holds. It remains to prove (\ref{rema}). Acting in the same way we
get
\begin{equation}\label{nach2}
\| F_{ij} (\psi_n - \phi_0 ) \|^2   = \| J_1 F_{ij} (\psi_n -
\phi_0 ) \|^2 + \sum_{a\geq 2} \| F_{ij} J_a (\psi_n - \phi_0 )
\|^2 .
\end{equation}
Again, the first term on the right--hand side of (\ref{nach2})
goes to zero because for $(i \in \mathfrak{C}_1 , \: j \in
\mathfrak{C}_2) $ the term $J_1 F_{ij} $ is relatively $H_0$
compact. For the other terms, we use that $F_{ij}$ is $H_0$
bounded with a relative bound zero, which gives us
\begin{equation}\label{nach3}
\| F_{ij} J_{a \geq 2}  (\psi_n - \phi_0 )  \| \leq b \| H_0 J_{a
\geq 2}  (\psi_n - \phi_0 ) \| + c\| J_{a \geq 2}  (\psi_n -
\phi_0 ) \|
\end{equation}
for some constants $b,c > 0$, where $b$ can be chosen arbitrary
small. Using the properties of $J_a $ it is easy to show that $J_{a \geq 2}
(\psi_n - \phi_0 )$ is a uniformly $H_0$-bounded
sequence since $(\psi_n - \phi_0 )$ is one (cf.
Lemma~\ref{forg}). Thus, the first term on the rhs in (\ref{nach3}) can
be made as small as pleased by choosing $b$ small, while the last
term goes to zero by (\ref{Ja}). Hence, indeed (\ref{rema}) is true,
which finally proves (a).

To prove (b) note that by (\ref{rema})
\begin{equation}\label{k112:}
\lim_{Z_n \to Z_{cr}} \bigl( (\psi_n-\phi_0), W(Z_n) (\psi_n-\phi_0) \bigr) = 0.
\end{equation}
Substituting (\ref{k112:}) into (\ref{zuzu}) we get
\begin{equation}\label{k114:}
\lim_{Z_n \to Z_{cr}} \Bigl( (\psi_n-\phi_0), \bigl[ H_{thr}(Z_n) - E_{thr}(Z_n)\bigr] (\psi_n-\phi_0) \Bigr) = 0.
\end{equation}
Inserting the identity $1 = P_{thr}(Z_n) + (1-P_{thr}(Z_n))$ into
(\ref{k114:}) and using (\ref{need1}) together with the fact that
$\left[H_{thr} - E_{thr}\right]P_{thr} = 0$ we finally prove (b). \end{proof}

\begin{lemma}\label{7} The following inequality holds
\begin{equation}\label{lem::7}
\sup_{k > 0} \left\| \Bigl[ p_{r}^2 + k^2 + \eta \Bigr]^{-1} P_{exp} \: \chi_{\{ \Omega \} } \right\| < \infty
\end{equation}
\end{lemma}

\begin{proof}
For the norm of the operator in (\ref{lem::7}) we have
\begin{gather}
\left\| \Bigl[ p_{r}^2 + k^2 + \eta \Bigr]^{-1} P_{exp} \: \chi_{\{ \Omega \} } \right\| = \\
\left\| \Bigl[ p_{r}^2 + k^2 + \eta \Bigr]^{-1} e^{-\frac a4 \left( \frac{r - C_0 }{C_1}\right)^{1/p} }P_{exp} \: e^{\frac a2 |\xi|}e^{-\frac a2 |\xi|} e^{\frac a4 \left( \frac{r - C_0 }{C_1}\right)^{1/p} } \chi_{\{ \Omega \} } \right\| \leq \\
\left\| \Bigl[ p_{r}^2 + k^2 + \eta \Bigr]^{-1} e^{-\frac a4 \left( \frac{r - C_0 }{C_1}\right)^{1/p} } \right\| \times \left\|  P_{exp} \: e^{\frac a2 |\xi|} \right\| \times \left\| e^{-\frac a2 |\xi|} e^{\frac a4 \left( \frac{r - C_0 }{C_1}\right)^{1/p} } \chi_{\{ \Omega \} } \right\|_{\infty}
\end{gather}

The first operator norm is uniformly bounded for all $k > 0$ by Lemma~\ref{expon}. The second norm is bounded by the definition of $P_{exp}$. And the third norm is bounded from the definition of the set $\Omega$.  \end{proof}

Let us make some additional remarks to Theorem~\ref{th2} and its proof. There is an apparent difficulty with including fermions into this approach because we require that $\phi_{thr} \geq 0$. This is a serious limitation for the present method. In the forthcoming article \cite{prl} we shall demonstrate how this difficulty can be circumvented in the case when the pair potentials are the sums of short--range and pure Coulomb terms. With an additional effort \cite{prl} one can lift the restriction saying that the lowest dissociation threshold corresponds to two particular clusters. This would open the way to inclusion into this approach identical particles (not fermions), for which the dissociation threshold might be degenerate due to symmetry.

\section{Behavior of Bound States in the Problem of three Coulomb
Charges}\label{sec:sec6}

Here we discuss a direct application of the above methods to the problem of stability of three Coulomb charges, for details see Refs.~ \cite{martin}.
This problem is particularly interesting
because it admits the coexistence of two types of behavior of bound
states as they approach the dissociation threshold, namely spreading and
non--spreading. Let us introduce the notations. There are three particles charges
$\{ q_1 , q_2, -1 \} $ and masses $\{ m_1 , m_2, m_3 \} $. The position
vectors are $x_i \in \mathbb{R}^3$ corresponding to the particle
mass $m_i$. The Hamiltonian \cite{jmpold} with a separated center of mass motion
in the Jacobi coordinates $\xi = x_3 - x_2$, $r = x_1 - x_2 - s\xi$,
reads
\begin{equation}\label{ap1}
    \tilde H = \frac{p_{\xi}^2}{2 \mu_{23}}   - \frac{q_2}{|\xi |} + \frac{p_{r}^2}{2 \mu}
- \frac{q_1}{|(1-s)\xi - r |} + \frac{q_1 q_2}{|a\xi +r|}
\end{equation}
where $\mu_{ik} = m_i m_k /(m_i + m_k )$, $\mu = m_1 (m_2 +m_3
)/(m_1 + m_2 + m_3 )$ are reduced masses and $s = m_3 / (m_3 + m_2)$.

We keep the masses fixed, which makes the Hamiltonian $\tilde H$ depend on parameters $(q_1 , q_2)$ alone. A typical stability diagram \cite{martin} for $\tilde H$ is sketched in Fig.~1, where the points in the shaded area represent the values of $(q_1 , q_2)$ for which $\tilde H$ is stable. Under \emph{stability} we mean that $\tilde H$ has at least one bound state below the bottom of the continuous spectrum. The properties of the stability diagram are discussed in detail in \cite{martin}. We mention some key features here. In the square $\{q_{1,2}|\: 0 < q_{1,2} <1\}$ the system is always stable (the physical reason is that at long distances there appears a Coulomb attraction between the bound pair and the third particle, which accommodates, in fact, infinitely many bound states). There are two possible dissociation thresholds formed by the pairs of charges $(-1, q_1)$ and $(-1, q_2)$ and the line of equal energy thresholds is given by the equation $\mu_{23} q_{2}^2 = \mu_{13} q_{1}^2 $. It is convenient to introduce the set $\mathfrak{D} = \{ q_1 , q_2 | q_1 >0 , \mu_{23} q_{2}^2 >
\mu_{13} q_{1}^2 \}$ such that for $(q_1 , q_2) \in \mathfrak{D}$ the
bottom of the continuous spectrum corresponds to the dissociation $(123) \to
(23) + 1$. The stability area is formed by two convex arcs, which form a cusp on the line of equal energy thresholds, just like in Fig.~1. Without loosing generality we concentrate our attention on the points within $\mathfrak{D}$. Other properties of the stability diagram are as follows. If the point $q^{(s)}_1, q^{(s)}_2 \in \mathfrak{D}$ is stable then all points in $\{q_{1,2}|\: q_1 \geq q^{(s)}_1, q_2 = q^{(s)}_2 \} \cap \mathfrak{D}$ and all points in $\{q_{1,2}|\: q_2 \leq q^{(s)}_2, q_1 = q^{(s)}_1 \} \cap \mathfrak{D}$ are also stable. There exists a constant $q_1^0$, $(q_1 = q_1^0, q_2 = 1) \in \mathfrak{D}$, such that all points on the peace of a line $\{q_{1,2}|\: 0 \leq q_1 \leq q^{0}_1, q_2 = 1 \} $ are unstable, while all points on $\{q_{1,2}|\: q_1 > q^{0}_1, q_2 = 1 \} \cap \mathfrak{D}$ are stable.
Thus the border of the stability domain in $\mathfrak{D}$ is formed by $\{q_{1,2}|\: 0 \leq q_1 \leq q^{0}_1, q_2 = 1 \} $ and by an arc, which starts from $(q_1 = q_1^0, q_2 = 1)$ and ends somewhere on the line of equal energy thresholds.

\begin{figure}
\begin{center}
\includegraphics[height=.2\textheight]{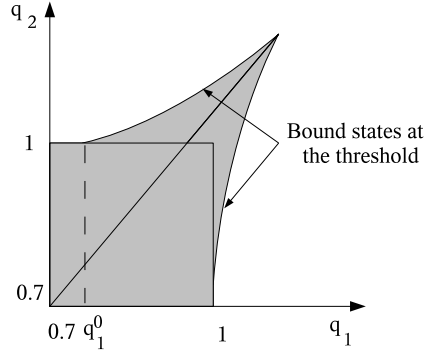}
\caption{Typical
stability diagram (sketch) for three Coulomb charges $\{ -1, q_1 ,
q_2 \} $, the shaded area representing stable systems. On the arcs
of stability curve where either $q_1
> 1$ or $q_2
> 1$ there are bound states at the threshold.}
\end{center}
\end{figure}

Using the results from the previous sections we can prove the following
\begin{theorem}\label{zaeb:}
Let $(q_1, q_2)\in \mathfrak{D}$ be a point on the stability border. Then: (a) if $q_2 >1 $ then at this point $\tilde H$ has a bound state at the threshold; (b) if   $q_1 < q_1^0$ and $q_2 = 1$ then at this point $\tilde H$ has no bound states at the threshold.
\end{theorem}
\begin{proof}
It makes sense to make a scaling transform of $\tilde H$. Let us rescale the masses $m_i \rightarrow 2 m_i / \mu_{23}$ and multiply all potentials by the factor $1/q_2$. Then we obtain the Hamiltonian in the form of Eq.~(\ref{goham})
\begin{equation}\label{goham:1}
H(Z) = H_{thr} + \frac{p^{2}_r }{2 \mu } + W (r, \xi, Z),
\end{equation}
where we define $Z = (q_1 , q_2)$ and
\begin{gather}
W(r, \xi, Z) = - \frac{q_1}{q_2 |(1-s)\xi - r |} + \frac{q_1}{|s \xi + r  |} \label{ap2} \\
H_{thr} = \frac {p_{\xi}^2}4  - \frac 1{|\xi|} \label{ap3}
\end{gather}
Due to the scaling invariance of Coulomb interactions $H$ has the
same stability diagram as $\tilde H$, and the wave functions within $\mathfrak{D}$ acquire a finite scaling factor \cite{martin}.
The advantage of transforming $\tilde H$ into $H$ is that within $\mathfrak{D}$ the bottom of the continuous spectrum of $H$ is
constant and equal to $E_{thr} = -1$.

Now suppose that the point $Z_{cr} = \left(q^{(a)}_1, q_2^{(a)}\right)$ lies on the stability border. By condition (a) of the theorem there must exist $\varepsilon >0$ such that a square
\begin{equation}\label{square:1}
    B_\varepsilon = \{q_{1,2}|\: |q_i - q^{(a)}_i| \leq \varepsilon \quad \mathrm{for} \quad i=1,2\}
\end{equation}
lies above the line of equal energy thresholds, $B_\varepsilon \subset \mathfrak{D}$ and $q^{(a)}_2 - \varepsilon = 1 + \delta q$, where $\delta q >0$ is another fixed constant. There exists a sequence $Z_n \in B_\varepsilon$ of stable points, which converges to $Z_{cr}$. Now it is straightforward to check if the Theorem~\ref{th2} applies to $(H(Z), \mathcal{Z})$. Let us check the requirements R1-6. The requirement R1 is fulfilled if we set $F(y) = (q_1^{(a)} + \varepsilon)/|y|$. The requirements R2-4 and R6 are obviously fulfilled. From the standard formula for the energy levels of a hydrogen atom it follows that the requirement R5 is fulfilled if we set
\begin{equation}\label{epsi::}
|\Delta \epsilon| = \min \left[ \frac 14 , \min_{q_{1,2} \in B_{\varepsilon}} \left\{\frac{\mu_{13}q_1^2}{\mu_{23}q_2^2} - 1 \right\}\right]
\end{equation}
Because $B_{\varepsilon}$ lies above the line of equal energy thresholds $|\Delta \epsilon|$ is positive. By a direct check one can verify that
\begin{equation}\label{nomoq::}
2\mu W(Z) \geq \frac 1{ |r|^2} \quad \mathrm{if} \quad |r| \geq \frac{2(2 + \delta q)}{\mu (q^{(a)}_1  - \varepsilon)\delta q} + \frac{4 + \delta q}{\delta q}|\xi|
\end{equation}
for all $Z \in B_{\varepsilon}$. Thus (\ref{W}) is fulfilled and Theorem~\ref{th2} applies, which proves (a).

The proof of (b) is trivial. Suppose that $H(q_1, q_2 = 1)$, where $q_1 < q^0_1$ has a bound state $\phi \in D(H_0)$ at the threshold $(\phi, [H+1] \phi) = 0$. Because $(\phi, p_r^2 \phi) > 0$ and $(\phi, [H_{thr}+1] \phi) \geq 0$ we must have $(\phi, W \phi) <0$. From (\ref{ap2}) it would follow that $(\phi, [H(q_1 + \epsilon, q_2 = 1)+1] \phi) <0 $, where $\epsilon > 0$ and $q_1 + \epsilon < q^0_1$. From the variational principle this would mean that $H(q_1 + \epsilon, q_2 = 1)$ is stable, which contradicts the definition of $q^0_1$.
 \end{proof}

Now, consider a sequence of stable points converging to some point on the stability curve $(q_1^{(a)}, q_2^{(a)})$. Theorem~\ref{zaeb:} tells us that if this point on the stability curve has $q_2^{(a)} >1$ then the wave functions corresponding to the points of the sequence do not spread. On the contrary, if $q_2^{(a)} =1$ and $q_1^{(a)} < q_1^0$ then the sequence of wave functions fully spreads. We do not know, however, whether at the point $(q_1^0, 1)$ there is a bound state at the threshold (our conjecture: yes, there is).

\section{Conclusions}\label{sec:sec7}

Using explicit bounds on the two--particle Green's functions we have
shown how the eigenvalue absorption works in the many--body case. It
has been proved that the long-range repulsion does not let bound
states spread and forces a bound state at the threshold. Though two
cases were considered, namely a two--cluster decay and a
multi--particle decay without clusters, both can be combined to show
that if, for example, the lowest dissociation threshold corresponds
to the decay into positively charged clusters (even more than two),
then the bound state should not spread and there would be a bound
state at the critical point.

There are apparent difficulties with including fermions into the
present approach. The reason is that in this case $H_{thr}$ must be
projected on the antisymmetric space and this limits the application
of Lemma~\ref{bond}. If the antisymmetry requirement are imposed only on permutations between two clusters and $\phi_{thr} \geq 0$ then the present approach still works. In the case of the Coulomb
interaction \cite{prl} one can still incorporate fermions into the present framework by using
multipole expansions.

\appendix
\section{Strictly Borromean Systems vs Borromean Systems}

The following theorem supplements Theorem~\ref{striborro} in the sense that the requirement on the system being ``strictly Borromean with a bound larger than $\varepsilon_0$'' in the condition of the theorem can be replaced through the requirement that the system is Borromean for all $Z \in \mathcal{Z}$. For that we have to pay the price of strengthening the requirement R2 as follows
\begin{list}{$\tilde {\text R}$\arabic{foo}}
{\usecounter{foo}
    \setlength{\rightmargin}{\leftmargin}}
\setcounter{foo}{1}
\item
$\forall f(x) \in C^\infty_0 (\mathbb{R}^{3N-3} ):$
\begin{gather}
\lim_{Z_k \to Z_{cr}} \Bigl\| \bigl[ \bigr(V_{ij}\bigl)_+ (Z_k) - (V_{ij}\bigl)_+(Z_{cr}) \bigr] f \Bigr\|  = 0 \label{whoknjski}\\
\lim_{Z_k \to Z_{cr}} \Bigl\| \bigl[ \bigr(V_{ij}\bigl)_- (Z_k) - (V_{ij}\bigl)_-(Z_{cr}) \bigr] f \Bigr\| = 0 \label{whoknjski:2}
\end{gather}
where $\{Z_k\} = \mathcal{Z}/Z_{cr}$ and $1 \leq i<j\leq N$.
\end{list}
It is clear that R2 follows from $\tilde {\text R}$2.

\begin{theorem}\label{bor-strbor}
Suppose that $(H(Z), \mathcal{Z})$ satisfies R1, $\tilde {\text R}$2, R3-4 and for all $Z \in \mathcal{Z}$ the Hamiltonian $H(Z)$ is Borromean. If the pair potentials satisfy the condition (\ref{many}) then there exists $\varepsilon >0$ independent of $Z$ such that $H(Z)$ is strictly Borromean with a bound larger than $\varepsilon$ for all $Z \in \mathcal{Z}$.
\end{theorem}
\begin{proof}

We shall first prove the statement that if  $H(Z_0)$ is Borromean for some fixed $Z_0 \in \mathcal{Z}$ then it must  be strictly Borromean with some positive bound. The proof is by induction on the number of particles. Let the statement hold for $N$ particles. We must prove that it also holds for $N+1$ particles. Because $Z$ is fixed we omit the explicit dependence on $Z$. Let $H_{N+1}$ be the Hamiltonian of a Borromean system of $N+1$ particles, which we assume by contradiction being not strictly Borromean. By the induction hypothesis there is $\varepsilon_0 >0$ such that all  $N$--particle subsystems of $H_{N+1}$ are strictly Borromean with a bound larger than $\varepsilon_0$.
We shall define by $\tilde H (\varepsilon)$ the transformed Hamiltonian $H$, where the Hamiltonian $\tilde H (\varepsilon)$  is the same as $H$ with the exception that all pair interactions in it take the form $\tilde V_{ij} (\varepsilon) = V_{ij} - \varepsilon \bigl( V_{ij}\bigr)_- $, where $V_{ij}$ denote pair interactions in $H$. For $n = 1,2,\ldots$  we define the sequence $\varepsilon_n := \varepsilon_0 /2n$. With a sequence defined like this all $N$--particle subsystems of $\tilde H_{N+1} (\varepsilon_n)$ are strictly Borromean with a bound larger than $\varepsilon_0 /2$.

Because  $H_{N+1}$  is not strictly Borromean it follows that there must exist a particular subsystem of $N$ particles with the Hamiltonian $H_N$ and functions $f_n \in D(H_N)$ such that $(f_n ,\tilde H_N (\varepsilon_n) f_n) < 0$ for all $n$. Since $\tilde H_{N} (\varepsilon_{n})$ is strictly Borromean, the HVZ theorem tells us that there exist normalized wave functions $\psi_n \in D(H_N)$ such that $\tilde H_N (\varepsilon_{n}) \psi_n = E_n \psi_n$, where $E_n < 0$.  It is straightforward to show that $E_n \to 0$. Indeed, we can rewrite the equation $E_n = (\psi_n , H_N(\varepsilon_n)\psi_n)$ in the form of the following inequality
\begin{equation}\label{zugzwang:}
    |E_n | \leq -(\psi_n , H_N(0) \psi_n) + \varepsilon_n \sum_{i<j} (\psi_n , F_{ij} \psi_n) \leq \varepsilon_n \sum_{i<j} \| F_{ij} \psi_n \| ,
\end{equation}
where we have used that $H_N(0) = H_N \geq 0$ since $H_{N+1}$ is Borromean. By the arguments of Lemma~\ref{forg}
 $\| F_{ij} \psi_n \|$ is uniformly bounded and, hence, $|E_n| \to 0$ because $\varepsilon_n \to 0$. Now we consider the Hamiltonian $(\tilde H_N (\varepsilon_{n}), \mathcal{\tilde Z})$, where $\mathcal{\tilde Z} = \{\varepsilon_n\}_{n=1}^\infty \cup \{0\}$, {\em i.e.} the role of the parameter sequence is played by $\varepsilon_{n}$ and the critical value is equal to zero. It is easy to check that all the requirements of Theorem~\ref{striborro} are fulfilled and thus $\tilde H_N (0) = H_N$ has a bound state with zero energy. But this contradicts $H_{N+1}$ being Borromean.

Note that the proof above applies to the case $N+1 =3$ without addressing the induction hypothesis because Theorem~\ref{striborro} is valid for a two-particle system without any further conditions on its subsystems. This observation completes the proof by induction.

Now we can prove the theorem. The proof is again by induction. Suppose that for $N$ particles the theorem is true. We must prove it for $N+1$. By the above statement $H_{N+1}(Z_k)$ is strictly Borromean with a bound $\varepsilon_k >0$. We must prove that the sequence $\varepsilon_k$ does not accumulate at zero. By induction hypothesis all $N$ particle subsystems of $H_{N+1}(Z)$ for all $Z \in \mathcal{D}$ are strictly Borromean with a bound larger than $\varepsilon_0 >0$. Let us choose a subsequence $\varepsilon_{k_s}$ such that $4 \varepsilon_{k_s} < \varepsilon_0$ and $\varepsilon_{k_s} \to 0$ for $s \to \infty$. On one hand, there must exist $H_N(Z)$, a subsystem of $H_{N+1}(Z)$ and $f_s \in D(H_N)$ such that $(f_s , \tilde H_N(Z_{k_s}, 2\varepsilon_{k_s}) f_s) <0$. On the other hand, the Hamiltonian $\tilde H_N(Z_{k_s}, 2\varepsilon_{k_s})$ is strictly Borromean with a bound larger that $\varepsilon_0/2$. This means that there are $\psi_s \in D(H_N)$ and $E_s < 0$ such that $\tilde H_N(Z_{k_s}, 2\varepsilon_{k_s}) \psi_s = E_s \psi_s$. Now we define the parameter sequence as $\tilde Z_s := (Z_{k_s}, 2\varepsilon_{k_s})$ and $\tilde Z_s \to \tilde Z_{cr} := (Z_{cr}, 0)$. It is straightforward to check that all requirements of Theorem~\ref{striborro} are fulfilled for $\tilde H_N(\tilde Z_s) $. The requirement R2 of is fulfilled through the requirement $\tilde {\text R}$2 of the present theorem. Hence, we conclude that $\tilde H_N(\tilde Z_{cr}) = H_N (Z_{cr}) $ has a bound state at zero energy. But this contradicts the condition of the theorem, which says that $H_{N+1}(Z)$ is Borromean for all $Z \in \mathcal{Z}$. The case $N+1 = 3$ is proved in the same way without addressing the induction hypothesis.  \end{proof}

\end{document}